\documentclass[preprint]{aastex}

\begin{document}

\title{Narrow-line Seyfert 1 Galaxies from the Sloan Digital Sky Survey Early Data Release}
\author{Rik J. Williams, Richard W. Pogge, and Smita Mathur}
\affil{Department of Astronomy, The Ohio State University, Columbus, OH 43210-1173, USA}

\begin{abstract}
We present a sample of 150 narrow-line Seyfert 1s (NLS1s) found within the 
Sloan Digital Sky
Survey Early Data Release (EDR), only two of which were previously identified
as such.  This substantially increases the known number of NLS1s, and provides
a basic method by which to identify many more with subsequent releases of SDSS
data.  With its large size and homogeneous, well-defined 
selection criteria, this sample will help alleviate two major problems which
have plauged NLS1 research in the past; namely, their relative rarity 
and significant differences in selection
algorithms between the known samples.  45 of these SDSS-selected NLS1s
are detected at energies of 0.1--2~keV
in the ROSAT All-Sky Survey (RASS), and are found to have ultrasoft X-ray 
spectra with photon
indices of $\Gamma \ga 2$, in agreement with previous results for NLS1s.  
However,
about 10--20 of those NLS1s that were not detected by ROSAT have 
optical properties very similar to the detected objects, and so should also have
been detected by the RASS.  This may be due to either significant
intrinsic absorption in many NLS1s, or a significant sub-class of NLS1s
that have uncharacteristic, intrinsically flatter (hence harder) X-ray
spectral energy distributions.

\end{abstract}


\section{Introduction}
Since their initial classification by \citet{ostpog85}
narrow-line Seyfert 1s (NLS1s) have gained noteriety as interestingly extreme
examples of active galactic nuclei (AGN).  They were initially defined by their
relatively narrow permitted emission lines \citep[FWHM $\la$ 
2000~km~s$^{-1}$;][]{goodrich89}, strong \ion{Fe}{2} relative to H$\beta$, 
and weak [\ion{O}{3}].  In their analysis of 87 bright AGN, \citet{bg92} found  
a strong anticorrelation between the strengths of the [\ion{O}{3}] and 
\ion{Fe}{2} lines (the primary correlation behind their so-called 
eigenvector 1 or Principal Component 1
[PC1]).  NLS1s lie at the extreme, 
low--[\ion{O}{3}] end of PC1.  The authors suggested that this may be due
to a high accretion rate relative to the Eddington rate.  A more recent
analysis by \citet{boroson02} reinforces this claim, noting that NLS1s
consistently exhibit the lowest estimated central black hole masses for 
similar luminosities and the highest inferred relative
accretion rates among the various AGN types.  

Much attention has been devoted in recent years to the X-ray spectra of NLS1s.
While AGN have long been known to emit a substantial fraction of their 
luminosity as X-rays, a soft X-ray ``excess'' was noted among
Seyfert galaxies \citep[e.g.,][and references therein]{puch92}.  
Boller, Brandt, \& Fink (1996) found
a strong anticorrelation between the X-ray photon index ($\Gamma$, where
$f_E \propto E^{-\Gamma}$) and FWHM(H$\beta$), with NLS1s having $\Gamma \ga
2.5$.  Consequently, selection on the basis of ultrasoft X-ray emission has
proven effective in the discovery of new NLS1s \citep{grupe00}.  
This soft excess is thought to be the high-energy (Wien) tail of
thermal emission from the inner accretion disk.  Since extremely high
temperatures are required to produce this emission, it again follows that
NLS1s may be powered by low-mass black holes at high relative
accretion rates \citep{pounds95, wang96}.  
 
To date, a combination of X-ray and optical selection and serendipity 
has resulted in the discovery of a large number of NLS1s
\citep[see][for a review]{pogge2000}: for example, the 
sample compiled by \citet[][hereafter VVG01]{vvg01} consists of 64
NLS1s with z~$<$~0.1, B~$<$~17.0, and $\delta > -25\degr$, while the {\it
Catalog of Quasars and Active Nuclei} \citep[10th edition:][]{vv01} lists
205 NLS1s among Seyferts and QSOs.  
While these provide a starting point
for studying the role of NLS1s among AGN phenomena, the sample size is 
relatively small and
quite heterogeneous due to the wide variety of selection criteria employed.  
Since many known NLS1s were first discovered in X-rays, it is difficult
to ascertain with confidence whether the extreme X-ray softness exhibited by 
most catalogued NLS1s is a fundamental property or a subtle selection effect.
Clearly, a large and homogeneous optically selected sample would be 
advantageous in resolving these issues.

Such a sample is now becoming available in the form of the Sloan Digital 
Sky Survey \cite[SDSS;][]{york2000}, in particular the Early Data Release 
\citep[EDR;][]{stoughton02}, 
released in mid-2001.  The EDR contains spectra
 of approxmiately 4000 quasars
\citep{schneider02} as well as a large number of Seyfert galaxies and 
other AGN.  Photometric data are measured in five bands
\citep[$u^\prime,g^\prime,r^\prime,i^\prime,$ and $z^\prime$;][]{fukugita96},
and criteria based on these bands are used to select QSO candidates
for spectroscopic follow-up, as described in \citet{richards02}.
Parameters such as redshifts, magnitudes and linewidths are
stored in a searchable database, with photometric properties in the 
``PhotoObj'' class and spectral properties in the ``SpecObj'' and 
``SpecLine'' classes.  This database also has built into it an
``ExternalCatalog'' class, which contains all EDR objects within 60\arcsec\ 
of objects catalogued in the ROSAT All-Sky Survey (RASS), as well as 
some ROSAT-measured properties of these objects (see section~\ref{xrayprop}). 

By submitting a query restricted to objects with narrow H$\beta$ emission 
lines and then
analyzing the resulting spectra, we have identified 150 NLS1s in the EDR,
only two of which have been previously classified as such.  In the following
section we discuss in more detail the selection criteria and subsequent
analysis.  The overall selection methods and optical properties are discussed,
as well as objects which were previously known and/or misidentified.  Finally,
we report on those objects which were also observed in RASS, and give possible
reasons why some were not detected when they should have been.

\section{Candidate selection from SDSS}
Using the SDSS Query Tool\footnote{Available from 
\url{http://archive.stsci.edu/sdss/software}}, we searched for 
spectroscopically-targeted 
objects which were flagged as
QSOs and which exhibited narrow H$\beta$ lines.
Velocity dispersions were estimated through the relation
\begin{equation}
\mathrm{FWHM}(\mathrm{H}\beta) =  \frac{2.35c\sigma}{\lambda_0(1+z)}
\end{equation}
where $\sigma$ is the data member ``sigma'' in the SDSS SpecLine class, 
denoting the value of $\sigma$ for a Gaussian curve fit to each H$\beta$
line.  To account for the possibility that the recorded dispersion estimates 
are subject to systematics, we initially relaxed the selection criteria, 
excluding only objects with FWHM(H$\beta$)~$>$~3000~km~s$^{-1}$.
The resulting 950
spectra were then visually inspected and measured to identify the NLS1s.  
Many of these spectra exhibited characteristics 
of Seyfert 1.5 and Seyfert 2 galaxies (strong [\ion{O}{3}] compared
to H$\beta$, no evidence of 
\ion{Fe}{2}, obvious broad components H$\beta$ and H$\alpha$, etc.), 
while others were too faint or noisy to classify.  All in all, this initial 
cut removed about half of the candidates from the sample.

\begin{figure}
\plotone{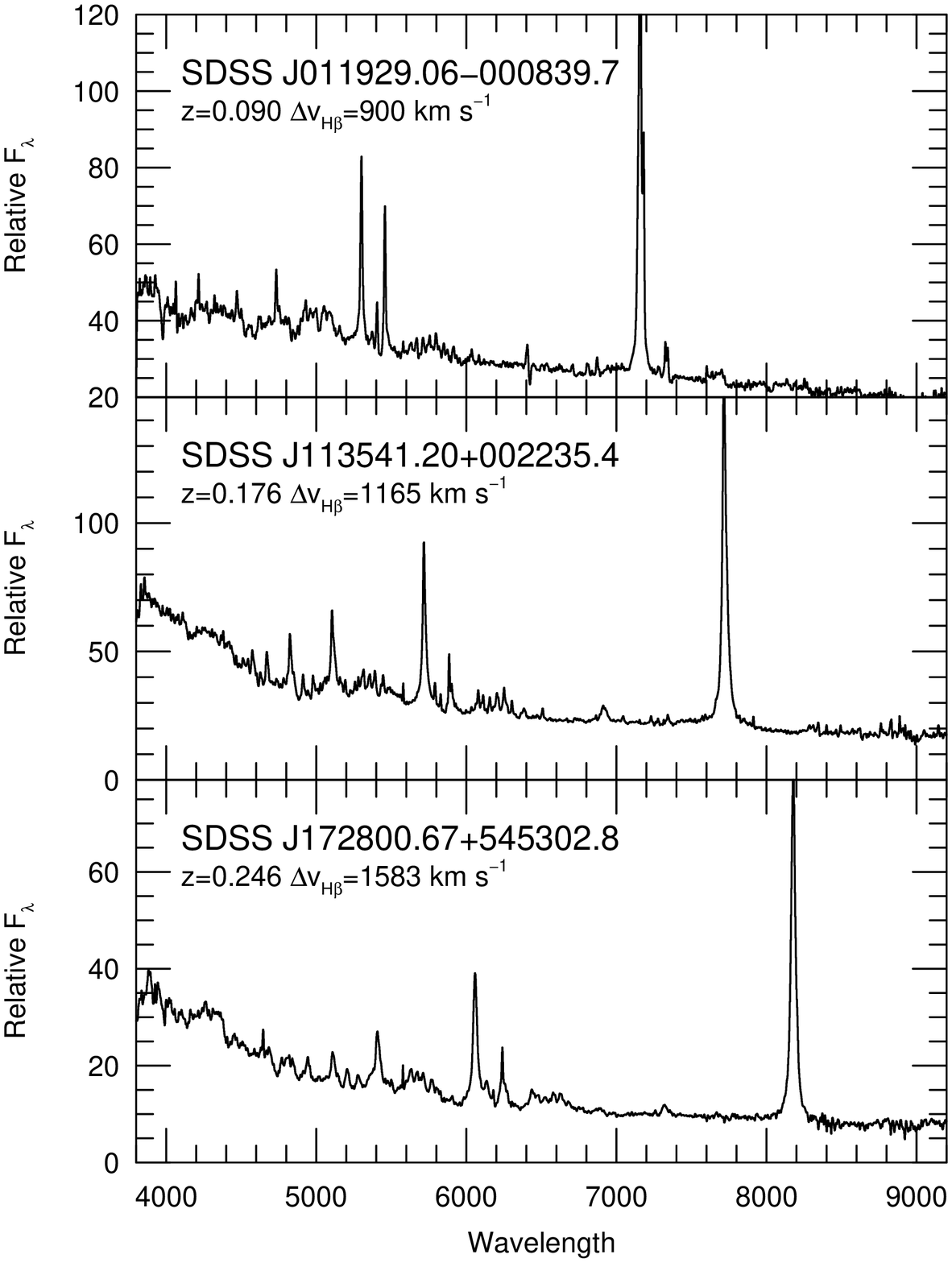}
\caption{Examples of three spectra from the NLS1 sample described in this
paper, taken from different FWHM(H$\beta$) and redshift regimes. \label{f1}}
\end{figure}

The remaining objects generally exhibit the combination of narrow H$\beta$, 
strong \ion{Fe}{2}
and weak [\ion{O}{3}] emission characteristic of NLS1s.  Each spectrum was 
first transformed from the \mbox{log$(\lambda)$} space used by SDSS onto
a linear wavelength scale and smoothed with
a three-pixel FWHM Gaussian filter.  The smoothing step is analogous to that 
used in the EDR presentation spectra returned as .GIF files.  We then 
performed a quadratic continuum fit near the H$\beta$ line, measured its
peak wavelength with a centroiding algorithm, and measured the width of the
line halfway between the fitted continuum and the line peak.  Since this 
method makes
no assumptions about the underlying emission-line profile, we took this to be  
an accurate measurement of FWHM(H$\beta$).  All objects which exhibited 
a velocity dispersion larger than 2000~km~s$^{-1}$, as well as those with 
evidence
of a weak very broad component in H$\beta$ or H$\alpha$ (when the latter was
visible in the SDSS spectral band), were removed.  The remaining 150 
objects thus 
satisfy the \citet{ostpog85} criteria and \citet{goodrich89} FWHM cutoff, 
and comprise the NLS1 catalog presented here (see Table~\ref{tbl}).  
Three of these NLS1 spectra are shown in figure~\ref{f1}, illustrating
the spectral shape and appearance for various redshift and FWHM(H$\beta$) 
regimes.

\begin{figure}
\plotone{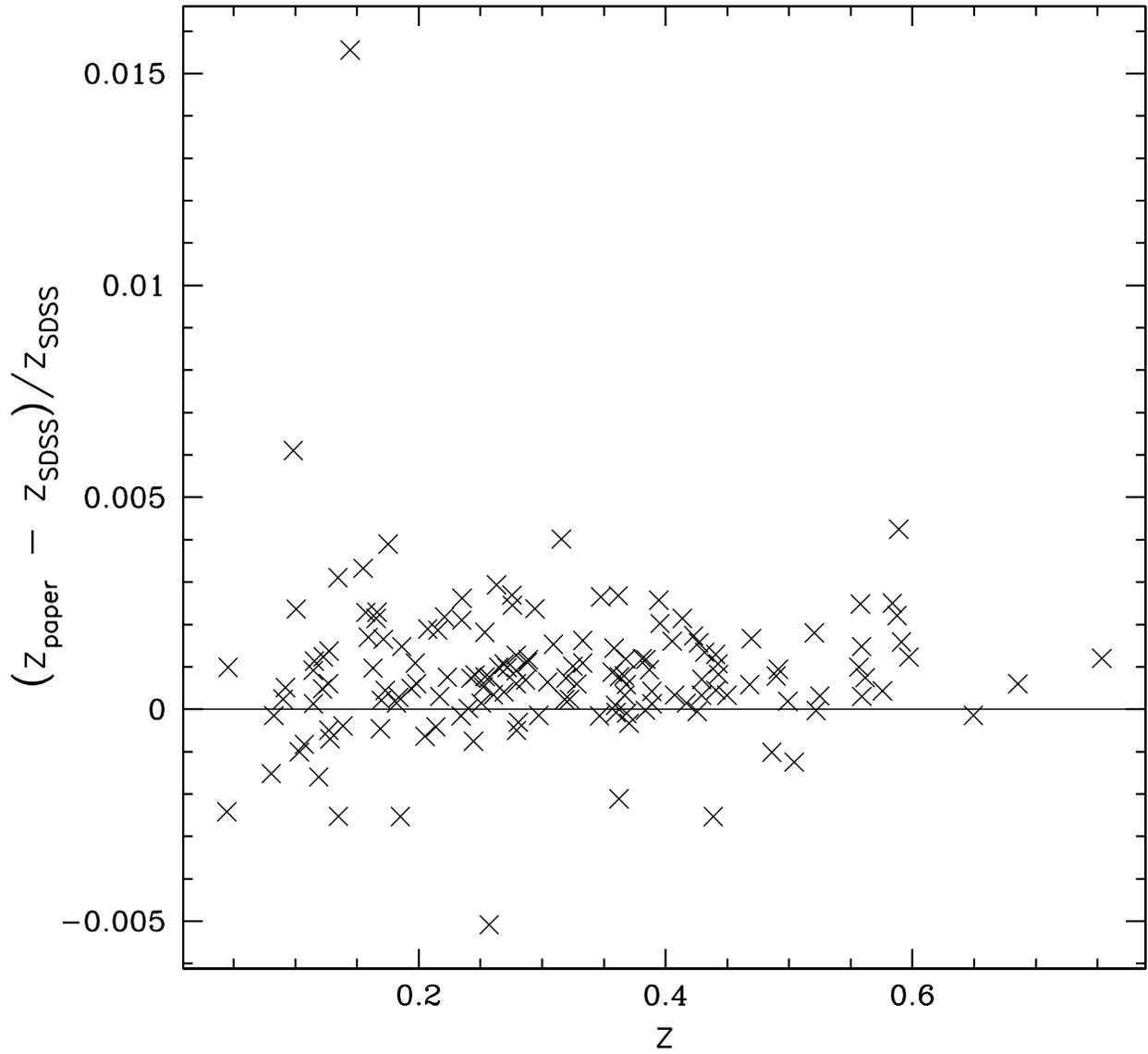}
\caption{Fractional differences between our redshift measurements and those made
by SDSS.  
\label{f2}}
\end{figure}

It is interesting to note the concordance between our measurements and those
reported in the SDSS EDR database; namely, those of the  redshift 
and H$\beta$ linewidth.
Figure~\ref{f2} shows excellent agreement between the two redshift
measurements with typical errors of 0.2\% or less.  Our redshift measurements
are systematically higher by about 0.1\%, but this is most likely due to 
properties of the H$\beta$ line itself since we base this redshift only 
on H$\beta$, rather than on the narrow forbidden lines that are probably
more representative of the systemic redshifts of the galaxies.  
When the most discrepant redshifts 
were re-measured using narrow forbidden lines (such as 
[\ion{O}{3}]\,$\lambda 5007$\AA), 
our redshifts fall much more in line with the SDSS measurements.  Since
SDSS bases their redshift measurement on a cross-correlation between many
lines and ours is only based on H$\beta$, we take the SDSS results
to be the more accurate and definitive.

\begin{figure}
\plotone{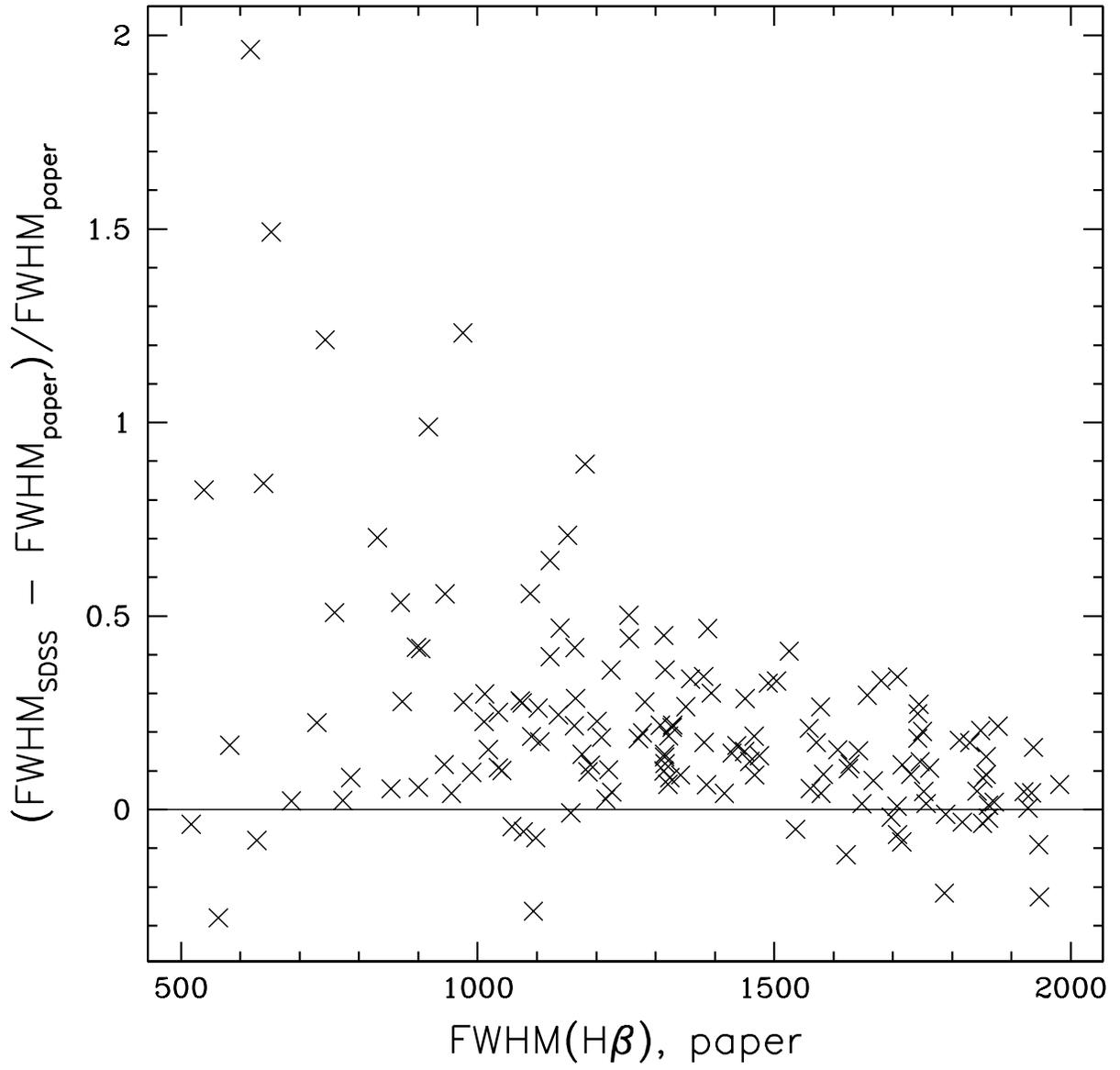}
\caption{Fractional differences between our FWHM measurements and those
made by SDSS.\label{f3}}
\end{figure}

On the other hand, 
there are large discrepancies between our FWHM measurements and those estimated
from the SDSS ``sigma'' parameter (see figure~\ref{f3}).  Not only 
is there a significant
amount of scatter, but our measurements are  systematically
lower than the SDSS estimates.  Most of the scatter is probably caused by the  
$\sim 2 \mathrm{\AA}$/pixel spectral resolution.  That is, if our FWHM 
measurements vary intrinsically by $\sim 2$~pixels, this would correspond to
an RMS variation 
of $\sim 250$~km~s$^{-1}$, which is very close to the observed scatter. 

The systematic offset is probably due to several factors.  First, such a 
discrepancy is not particularly surprising since the SDSS FWHM 
is based on a Gaussian fit to a profile better represented by a Lorentzian 
or more complicated shape (see, for example, VVG01).  These are 
compounded
by the proximity of the \ion{Fe}{2} complexes on either side of H$\beta$, 
which tends to drive the automatically-fitted continuum used by the
SDSS analysis pipeline higher depending on 
the \ion{Fe}{2} strength.  In some cases, the line-fitting algorithm
employed by SDSS appears to select only a broad component in the H$\beta$ 
line, again giving larger FWHM values than our measurements.  
Thus, while SDSS measurements are
quite useful for initial linewidth-based selection, careful follow-up 
measurements are absolutely necessary to take into account peculiarities  
in the individual spectra.

\section{Sample properties}

Table~\ref{tbl} lists the 150 objects which comprise this sample, 
along with various measured parameters and previous references to the
catalogued objects, found through a NASA/IPAC Extragalactic Database (NED) 
query.  
Although 48 objects had been picked up in surveys such as 2dF,
2MASS, LBQS, and various other projects, most had not been formally classified 
as NLS1s.  Of these, only two have
been previously identified as NLS1s: SDSS J014644.82--004043.2 (VVG01),
and SDSS J010226.31--003904.6 \citep{vv01}.  
\citet{puch92} list another,
SDSS J011703.58+000027.4 (also known as E0114-002),
as a Seyfert 1 with FWHM(H$\beta$)\,$=$\,2980~km~s$^{-1}$, 
substantially higher than our measurement of
975~km~s$^{-1}$; however, the authors mention that this object's H$\beta$ 
line may be contaminated.  Such contamination was not evident in the
SDSS spectrum, so we have included this object in our NLS1 sample.
The remaining objects are listed in NED under such generic 
labels as ``QSO,'' ``AGN,'' or ``Seyfert.'' A few are flagged as ``Sy1,'' but
it is unclear whether these were classified before \citet{ostpog85} first 
defined 
the NLS1 class.  In the interest of brevity, Table~\ref{tbl} contains
special references only to
those objects flagged as some type of Seyfert galaxy in NED.
 
All in all, this sample represents 150 spectroscopically-selected NLS1s.
While it significantly increases the number of known NLS1s, the homogeneous 
sample selection criteria (through use of the SDSS EDR catalog) make it 
particularly useful for studies of the overall NLS1 population.  For $z<0.5$ 
(where most of our objects lie), there are 135 objects in our sample out of 
944 flagged as QSOs in the EDR.  If we take this to be
representative of the overall quasar population, this would imply that 
NLS1s make up roughly 15\% of AGN at low redshifts.  Until a more definitive
selection of both NLS1s and QSOs is made from the SDSS, this number
is only a rough approximation.  However, this is comparable to
the number quoted by \citet{osterbrock87}.

\subsection{Constraints}
The NLS1s we have found in the SDSS EDR span a range of redshifts from 
0.04 to 0.75, the upper bound
set by the requirement that [\ion{O}{3}]\,$\lambda 5007$\AA\ not exceed the 
SDSS spectrograph limit of $\sim 9200$\AA\  
(though in practice
only three objects have $z > 0.6$).  Although \citet{schneider02} warn that
the quasar selection criteria are inhomogeneous for the EDR, the 
criteria for objects at these lower redshifts are actually quite well defined
\citep{richards02}.  Since NLS1s are identified primarily by the H$\beta$
velocity dispersion and the [\ion{O}{3}] and \ion{Fe}{2} lines surrounding 
H$\beta$, we did not
attempt to find NLS1-like AGN at redshifts beyond the range where
H$\beta$ is seen.  Furthermore, a search for narrow-lined objects in the SDSS
Galaxy database yielded over 20,000 candidates, and thus it was not
feasible to consider objects in that database with the  
selection methods described in this paper.  By restricting
ourselves to objects flagged as QSO, our sample also falls within the SDSS
QSO color selection criteria described in \citet{richards02}, making the basic
selection very well understood.  Even with
these restrictions, there are enough objects to undertake a study of
characteristics of this NLS1 sample.

\subsection{X-ray properties\label{xrayprop}}

When these 150 objects are cross-referenced with the ROSAT All-Sky Survey
catalogue 
(using an SDSS ExternalCatalog query), we find that 52 lie within
60\arcsec\ of X-ray detected sources.
Power-law slopes were estimated using the ROSAT Hardness Ratio 1 parameter
(HR1), which is defined as \citep{rass}:
\begin{equation}
\mathrm{HR1} = \frac{\mathrm{B}-\mathrm{A}}{\mathrm{B}+\mathrm{A}}
\end{equation}
Here, A and B denote the number of counts in the 0.1--0.4 and 0.5--2.0~keV
bands, respectively.  Note that if there are zero counts in band A or B, 
HR1 becomes +1 or --1 respectively.  By using HR1 and the sensitivity curve
of ROSAT (as implemented in the PIMMS program\footnote{Portable, Interactive
Multi--Mission Simulator v3.2d, from NASA's High Energy Astrophysics Science
Archive Research Center, currently available at 
\url{http://heasarc.gsfc.nasa.gov/docs/software/tools/pimms.html}}) and 
Galactic $N_H$ obtained from the ``nh'' utility\footnote{Part of the FTOOLS 
package, available at
\url{http://heasarc.gsfc.nasa.gov/docs/software/ftools/}}, it
is possible to estimate the X-ray photon indices of ROSAT-detected
NLS1s.

\begin{figure}
\plotone{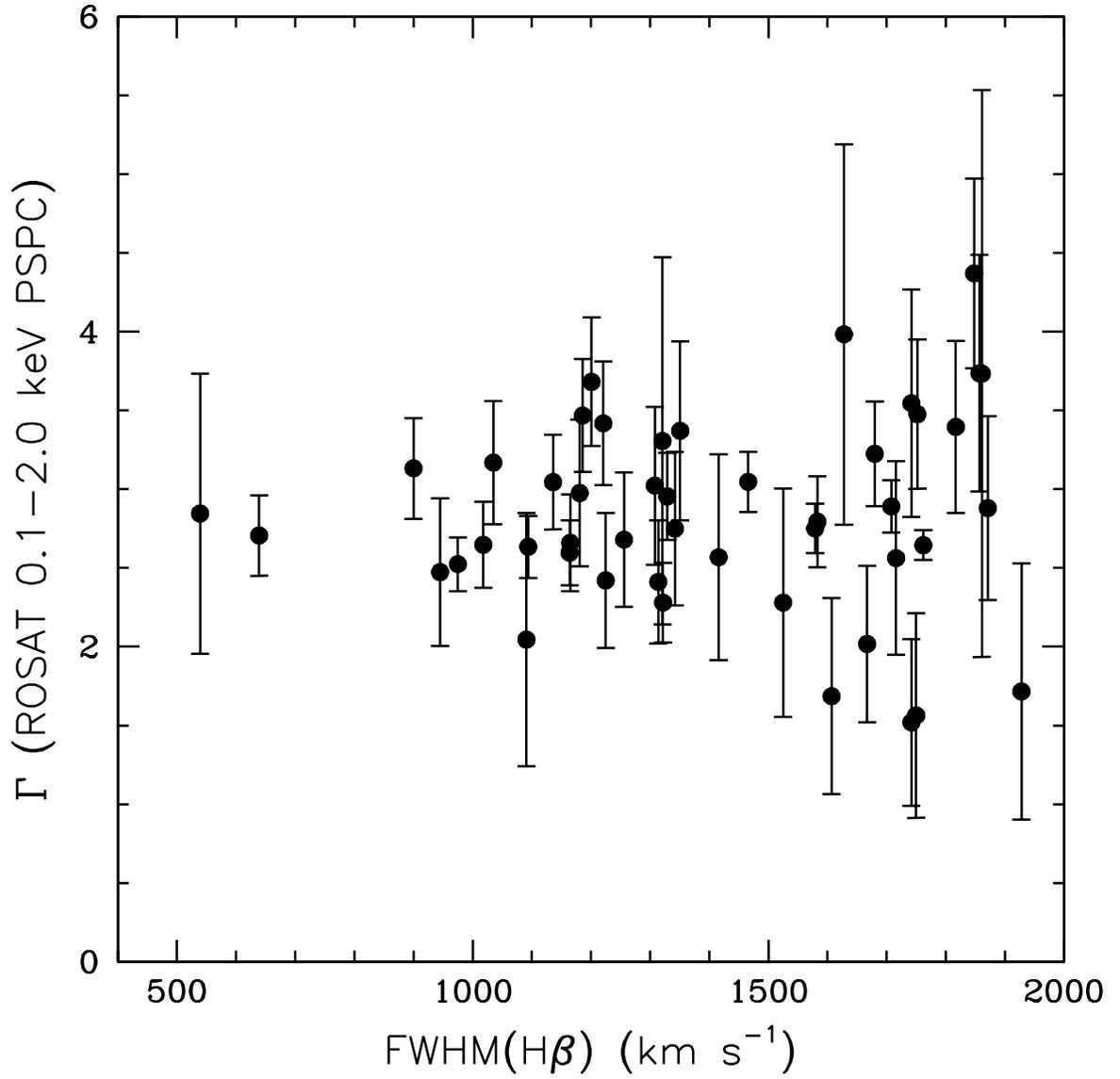}
\caption{Photon index vs. H$\beta$ velocity dispersion for the ROSAT-detected
NLS1s.\label{f4}}
\end{figure}

Of the 52 X-ray detected NLS1s in our sample, 7 have hardness ratios equal 
to (or within 
$1\sigma$ of) +1 or --1, with two in the former group and five in the latter.  
This may be indicitave of extremely hard or soft X-ray spectra, respectively.
However, these seven objects are all fairly faint in X-rays ($\la$ 0.04
counts~s$^{-1}$) and the two with $\mathrm{HR1} = +1$ have unusually high
Galactic \ion{H}{1} column densities; thus, it is unlikely that these
seven objects have unusually hard or soft spectral energy distributions.  
Power-law slopes were estimated with PIMMS for the 
remaining 45~NLS1s and plotted against FWHM(H$\beta$)
(see figure~\ref{f4}).
As in \citet{bbf96}, the photon indices of this sample of NLS1s span a range
from approximately $2 \la \Gamma \la 4.5$ with little or no apparent 
dependence on the H$\beta$ linewidth.
In particular, we note that
the vast majority of these sources exhibit ultrasoft spectra with $\Gamma > 2$,
as was seen in previous samples of soft X-ray selected NLS1s 
\citep[e.g.][for a review]{boller00}. 

\begin{figure}
\plotone{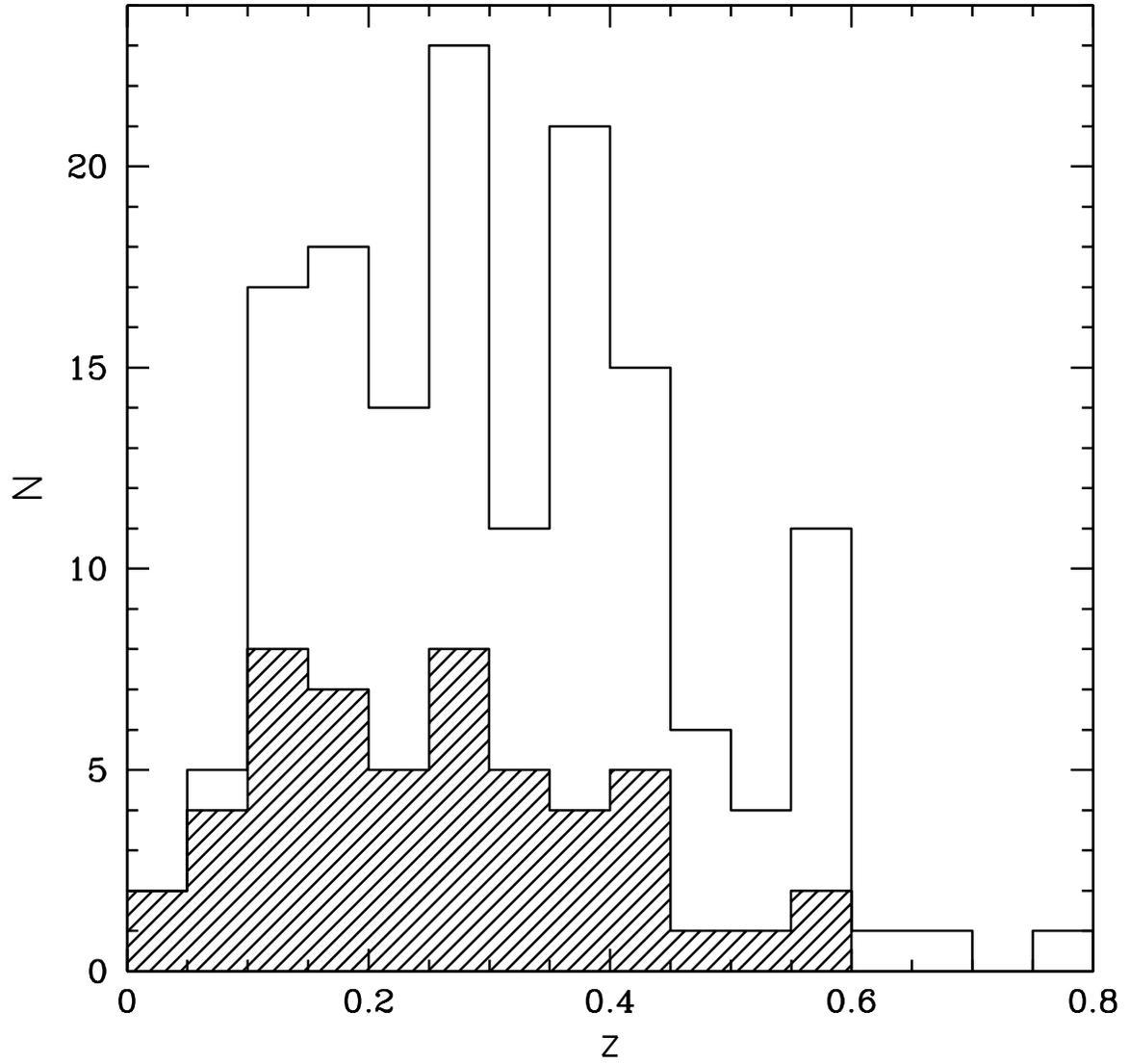}
\caption{Redshift distribution for this NLS1 sample.
ROSAT-detected NLS1s are shown as the shaded portion of the histogram.
\label{f5}}
\end{figure}

\begin{figure}
\plotone{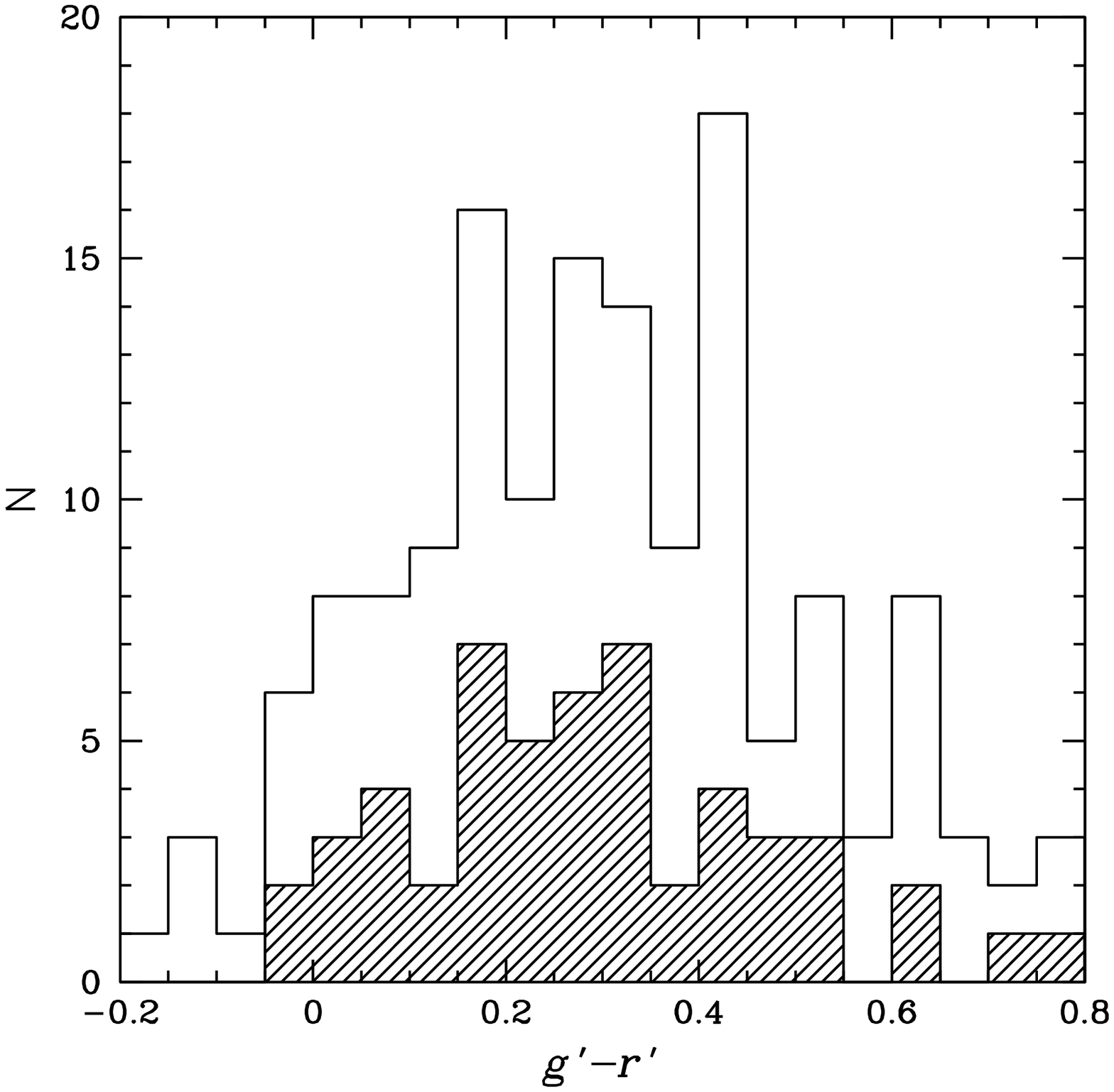}
\caption{SDSS $g^\prime-r^\prime$ color distribution for this NLS1 sample.
ROSAT-detected NLS1s are shown as the shaded portion of the histogram.
\label{f6}}
\end{figure}

\begin{figure}
\plotone{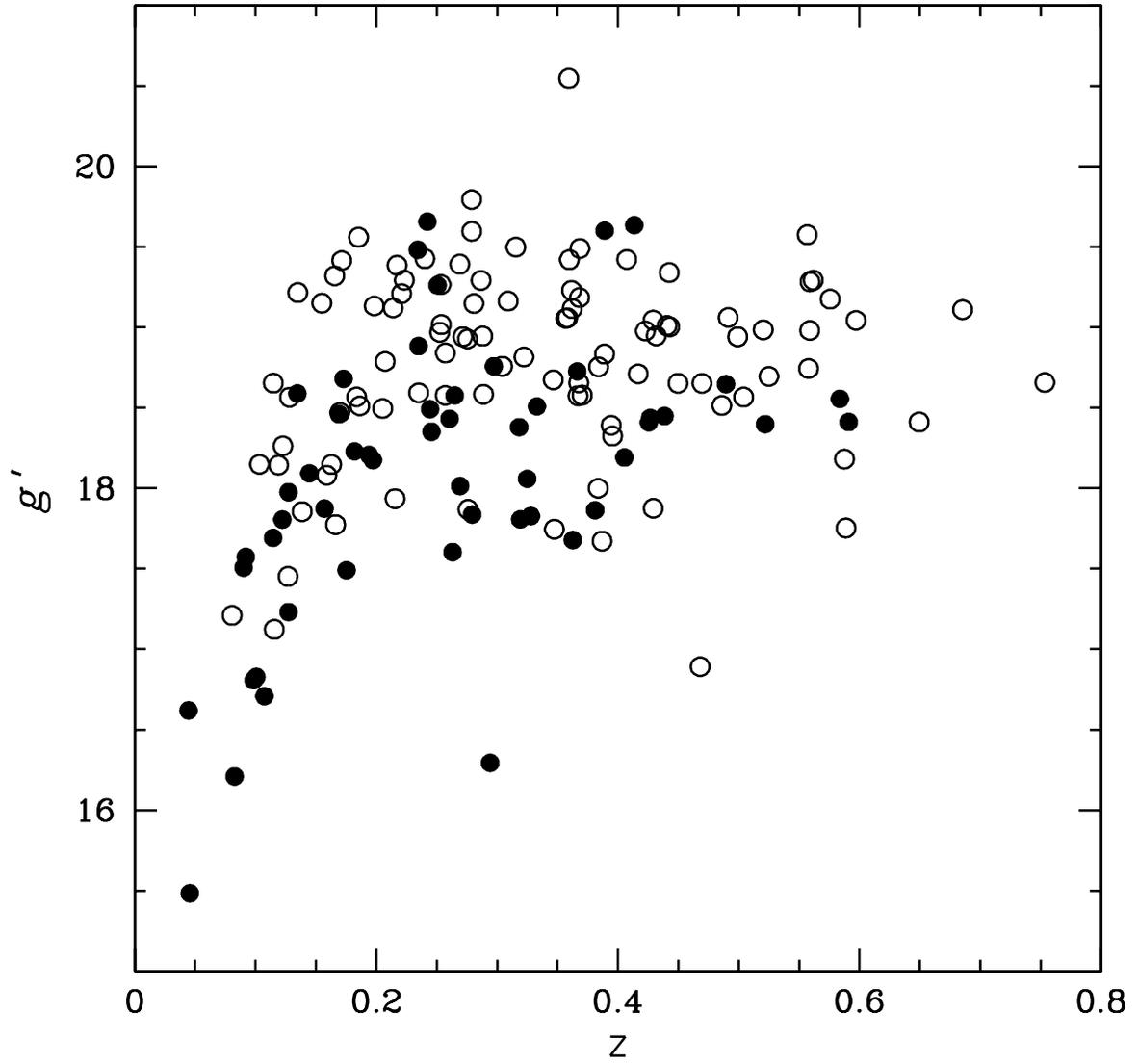}
\caption{Magnitude-redshift relation for NLS1s with Galactic 
$N_H < 4\times10^{20}$~cm$^{-2}$.
Filled circles denote those objects detected in the ROSAT All-Sky Survey,
while open circles were not detected.\label{f7}}
\end{figure}

Redshift and $g^\prime - r^\prime$ color distributions are shown in 
figures~\ref{f5} and~\ref{f6} respectively, with ROSAT-detected objects 
shown as the shaded
portion of each histogram.
Most of these RASS sources are optically bright ($g^\prime\la 18.5)$;
additionally, there appears to be a slight bias toward sources with lower
redshifts and Galactic \ion{H}{1} column densities.  Even with this taken into 
account by excluding objects with 
$N_H \geq 4\times10^{20}\ \mathrm{cm}^{-2}$,
we still see several optically-bright, low redshift sources which should have
been easily detected by RASS assuming similar spectral energy distributions, but
which were not (see figure~\ref{f7}).  

There are two likely interpretations for this.  The first is that these
undetected sources could have X-ray properties similar to the detected
sources, but the soft X-rays are suppressed by a large
($N_H \ga 10^{21}$\ cm$^{-2}$) intrinsic column density.  If this 
were the case, we would expect to see a pronounced difference in the
spectral continuum shapes and/or the photometric colors between the
objects that are detected in X-ray and those that are not; however, 
no such reddening is seen.
The soft X-ray weak objects could be members of a gas-rich and dust-poor
population \citep[as proposed by][]{risaliti01}, which would account for
the apparent lack of visual extinction, but this seems somewhat contrived.

The second possibility is that the undetected objects represent a new
population of NLS1s with intrinsically flatter (and hence harder) X-ray 
slopes.  In this
case, the lack of soft X-rays would be due to a fundamental difference
in the central black hole and accretion disk properties.  For example, if
these objects had larger black hole masses and high accretion rates, 
the accretion disk spectral energy distribution would shift to lower energies,
effectively flattening out the 0.1--2~keV spectrum.  Assuming similar 
spectral properties, these NLS1s would not have been detected in the RASS.  

It should also be noted that strong X-ray variability could
result in a significant fraction of NLS1s not detected by RASS.  However, it
is unlikely that such a large number of bright NLS1s would exhibit this
degree of variability, and would coincidentally be X-ray faint at the
time of observation by RASS.  Whichever interpretation
is correct, the X-ray undetected NLS1s are possibly something new and
intriguing within NLS1 phenomena.  Further
X-ray observations of these NLS1s over a larger energy range
with higher sensitivity may help to decipher the underlying cause of 
this observation.

\section{Conclusions\label{conclusions}}

The 150 SDSS-selected NLS1s presented in this paper represent a significant 
increase
in the total known number of these extreme AGN.  They comprise approximately
15\% of the EDR ``QSO'' database at $z \la 0.5$ and have very well-defined
color and linewidth selection criteria.  45 of these NLS1s were also 
detected with good confidence in the 0.1--2~keV band of the ROSAT 
All-Sky Survey and exhibit 
ultrasoft X-ray spectra.  Of the NLS1s that were \emph{not} detected, several
have similar optical properties to NLS1s seen in the RASS, and thus should have
been detected as well.  This may be due to either high intrinsic
absorption or harder X-ray spectra (or both) among the undetected objects.  
More optical and X-ray data will almost certainly be helpful in determining
the cause behind this.

It should also be noted that while most of our objects clearly fall within
the defining NLS1 criteria, a substantial fraction are near the
2000~km~s$^{-1}$ FWHM cutoff.  Additionally, several of the spectra had low
signal--to--noise ratios.  This may hide characteristics (such as a broad
component in H$\beta$) which would reclassify the object as a Seyfert 1.5
or other type.  While these objects all appear to be NLS1s from the data 
given, it is likely that some may be reclassified when
higher resolution, higher signal-to-noise spectra are obtained.  Thus, 
this sample should be considered a list of very strong NLS1 candidates
rather than a definitive list.  Nevertheless, this sample demonstrates that
large numbers of NLS1s (and other interesting objects) with well-constrained
selection criteria can indeed be
found in the SDSS Early Data Release.
Since the EDR
represents only about 5\% of the total spectroscopic survey
\citep{stoughton02}, the full SDSS catalog will provide a definitive
resource for the discovery and study of NLS1s.

\acknowledgements
The authors would like to thank David Weinberg for helpful discussions.  
We also thank the anonymous referee for helpful comments.
RJW acknowledges the support of an Ohio State University Fellowship.
SM is supported by Chandra X-ray Observatory grant G01-2118X from
Smithsonian Astrophysical Observatory.
Funding for the creation and distribution of the SDSS Archive has been 
provided by the Alfred P. Sloan Foundation, the Participating Institutions, 
the National Aeronautics and Space Administration, the National Science 
Foundation, the U.S. Department of Energy, the Japanese Monbukagakusho, 
and the Max Planck Society. The SDSS Web site is \url{http://www.sdss.org/}.
The SDSS is managed by the Astrophysical Research Consortium for the 
Participating Institutions. The Participating Institutions are The 
University of Chicago, Fermilab, the Institute for Advanced Study, the 
Japan Participation Group, The Johns Hopkins University, Los Alamos 
National Laboratory, the Max-Planck-Institute for Astronomy, the 
Max-Planck-Institute for Astrophysics, New Mexico State University, 
Princeton University, the United States Naval Observatory, and the 
University of Washington.
This research has made use of the NASA/IPAC Extragalactic Database (NED) 
which is operated by the Jet Propulsion Laboratory, California Institute 
of Technology, under contract with the National Aeronautics and Space 
Administration.

\clearpage

\clearpage

\begin{deluxetable}{lccrrr}
\tablecaption{NLS1s from the SDSS Early Data Release\label{tbl}}
\tablehead{
\colhead{SDSS Name\tablenotemark{a}} & \colhead{z} & \colhead{FWHM(H$\beta$)} &
\colhead{HR1} & \colhead{$\Gamma$} & \colhead{Notes}
}

\startdata
J000109.14-004121.5 &0.417 &1209 &\nodata &\nodata &\nodata\\
J000834.72+003156.2 &0.263 &1351 &$-0.46 \pm 0.29$ &$3.4 \pm 0.6$ &\nodata\\
J001327.31+005232.0 &0.363 &1742 &$-0.57 \pm 0.31$ &$3.5 \pm 0.7$ &\nodata\\
J002213.00-004832.7 &0.214 &1429 &\nodata &\nodata &\nodata\\
J002233.27-003448.6 &0.504 &1388 &\nodata &\nodata &\nodata\\
J002305.03-010743.5 &0.166 &1157 &\nodata &\nodata &\nodata\\
J002752.39+002615.8 &0.205 &1830 &\nodata &\nodata &\nodata\\
J003024.94+000254.5 &0.288 &743 &\nodata &\nodata &\nodata\\
J003238.20-010035.2 &0.092 &639 &$-0.10 \pm 0.17$ &$2.7 \pm 0.3$ &\nodata\\
J003431.74-001312.7 &0.381 &1314 &$0.03 \pm 0.27$ &$2.4 \pm 0.4$ &\nodata\\
J003711.00+002128.0 &0.235 &617 &\nodata &\nodata &\nodata\\
J004052.14+000057.3 &0.405 &1278 &\nodata &\nodata &\nodata\\
J004338.54-005814.7 &0.559 &1122 &\nodata &\nodata &\nodata\\
J005446.16+004204.1 &0.234 &1225 &$0.09 \pm 0.29$ &$2.4 \pm 0.4$ &\nodata\\
J005921.37+004108.9 &0.423 &1625 &\nodata &\nodata &\nodata\\
J010226.31-003904.6 &0.294 &1680 &$-0.24 \pm 0.20$ &$3.2 \pm 0.3$ &1\\
J011357.93-011139.8 &0.754 &1842 &\nodata &\nodata &\nodata\\
J011703.58+000027.4 &0.046 &975 &$0.20 \pm 0.11$ &$2.5 \pm 0.2$ &2\\
J011712.81-005817.5 &0.486 &1937 &\nodata &\nodata &\nodata\\
J011929.06-000839.7 &0.090 &900 &$-0.22 \pm 0.20$ &$3.1 \pm 0.3$ &3\\
J013046.16-000800.8 &0.253 &1648 &\nodata &\nodata &\nodata\\
J013521.68-004402.2 &0.098 &1181 &$-0.23 \pm 0.29$ &$3.0 \pm 0.5$ &\nodata\\
J013842.05+004020.0 &0.520 &1035 &\nodata &\nodata &\nodata\\
J013940.99-010944.4 &0.194 &1091 &$0.36 \pm 0.49$ &$2.0 \pm 0.8$ &\nodata\\
J014234.41-011417.4 &0.244 &1607 &$0.55 \pm 0.31$ &$1.7 \pm 0.6$ &\nodata\\
J014412.77-000610.5 &0.359 &1041 &\nodata &\nodata &\nodata\\
J014542.78+005314.9 &0.389 &1255 &\nodata &\nodata &\nodata\\
J014559.45+003524.7 &0.166 &1075 &\nodata &\nodata &\nodata\\
J014644.82-004043.2 &0.083 &1164 &$0.00 \pm 0.14$ &$2.6 \pm 0.2$ &4\\
J014951.66+002536.5 &0.252 &563 &\nodata &\nodata &\nodata\\
J015249.76+002314.7 &0.589 &1852 &\nodata &\nodata &\nodata\\
J015652.43-001222.0 &0.163 &1324 &\nodata &\nodata &\nodata\\
J020431.64+002400.5 &0.171 &1077 &\nodata &\nodata &\nodata\\
J021610.56+000538.4 &0.384 &1467 &\nodata &\nodata &\nodata\\
J021652.47-002335.3 &0.304 &854 &\nodata &\nodata &\nodata\\
J022205.37-004948.0 &0.525 &1571 &\nodata &\nodata &\nodata\\
J022756.28+005733.1 &0.128 &773 &\nodata &\nodata &\nodata\\
J022841.48+005208.6 &0.186 &990 &\nodata &\nodata &\nodata\\
J022923.43-000047.9 &0.558 &1386 &\nodata &\nodata &\nodata\\
J023057.39-010033.7 &0.649 &1947 &\nodata &\nodata &\nodata\\
J023211.83+000802.4 &0.432 &1746 &\nodata &\nodata &\nodata\\
J023414.58+005707.9 &0.269 &1381 &\nodata &\nodata &\nodata\\
J024037.89+001118.9 &0.470 &1789 &\nodata &\nodata &\nodata\\
J024651.91-005931.0 &0.468 &1504 &\nodata &\nodata &\nodata\\
J025501.19+001745.5 &0.360 &904 &\nodata &\nodata &\nodata\\
J030031.31+005357.2 &0.198 &1536 &\nodata &\nodata &5\\
J030417.78+002827.4 &0.044 &1321 &$0.46 \pm 0.39$ &$3.3 \pm 0.8$ &\nodata\\
J030639.57+000343.2 &0.107 &1525 &$0.80 \pm 0.17$ &$2.3 \pm 0.7$ &\nodata\\
J031427.47-011152.4 &0.387 &1812 &\nodata &\nodata &\nodata\\
J031542.64+001228.7 &0.207 &870 &\nodata &\nodata &\nodata\\
J031630.79-010303.6 &0.368 &1226 &\nodata &\nodata &\nodata\\
J032255.49+001859.9 &0.384 &1621 &\nodata &\nodata &\nodata\\
J032337.65+003555.7 &0.215 &1490 &\nodata &\nodata &\nodata\\
J032606.75+011429.9 &0.127 &686 &\nodata &\nodata &\nodata\\
J033027.21+005433.7 &0.443 &1315 &\nodata &\nodata &\nodata\\
J033059.06+010952.1 &0.557 &1946 &\nodata &\nodata &\nodata\\
J033429.44+000611.0 &0.347 &1316 &\nodata &\nodata &\nodata\\
J033854.25+005339.7 &0.279 &1314 &\nodata &\nodata &\nodata\\
J033923.66-002310.3 &0.369 &1437 &\nodata &\nodata &\nodata\\
J034131.95-000933.0 &0.223 &897 &\nodata &\nodata &\nodata\\
J034326.51+003915.2 &0.499 &1315 &\nodata &\nodata &\nodata\\
J034430.03-005842.7 &0.287 &786 &\nodata &\nodata &\nodata\\
J094857.33+002225.5 &0.584 &1342 &$0.31 \pm 0.27$ &$2.8 \pm 0.5$ &\nodata\\
J095859.80+004718.9 &0.235 &1190 &\nodata &\nodata &\nodata\\
J100405.00-003253.4 &0.289 &582 &\nodata &\nodata &\nodata\\
J101314.86-005233.5 &0.276 &1578 &\nodata &\nodata &\nodata\\
J102059.72+010034.3 &0.588 &1715 &\nodata &\nodata &\nodata\\
J102450.52-002102.4 &0.322 &1382 &\nodata &\nodata &\nodata\\
J103031.41-001902.6 &0.562 &1787 &\nodata &\nodata &\nodata\\
J103222.58-000345.6 &0.559 &1707 &\nodata &\nodata &\nodata\\
J103457.29-010209.0 &0.328 &1394 &\nodata &\nodata &\nodata\\
J104132.35-003512.2 &0.135 &1316 &\nodata &\nodata &\nodata\\
J104210.03-001814.7 &0.115 &628 &\nodata &\nodata &\nodata\\
J104230.14+010223.7 &0.116 &1012 &\nodata &\nodata &\nodata\\
J104331.51-010732.9 &0.362 &1756 &\nodata &\nodata &\nodata\\
J104449.28+000301.2 &0.443 &1176 &\nodata &\nodata &\nodata\\
J105932.52-004354.7 &0.155 &1451 &\nodata &\nodata &\nodata\\
J110312.83+000012.5 &0.276 &1450 &\nodata &\nodata &\nodata\\
J111022.39-005544.5 &0.257 &1934 &\nodata &\nodata &\nodata\\
J111307.73+003210.4 &0.346 &976 &\nodata &\nodata &\nodata\\
J113102.28-010122.0 &0.242 &1928 &$0.60 \pm 0.37$ &$1.7 \pm 0.8$ &\nodata\\
J113541.20+002235.4 &0.175 &1165 &$-0.16 \pm 0.20$ &$2.7 \pm 0.3$ &3\\
J115023.59+000839.1 &0.127 &1136 &$-0.45 \pm 0.16$ &$3.0 \pm 0.3$ &3\\
J115306.95-004512.7 &0.357 &1102 &\nodata &\nodata &\nodata\\
J115412.77+010133.4 &0.490 &945 &$-0.11 \pm 0.32$ &$2.5 \pm 0.5$ &\nodata\\
J115533.50+010730.6 &0.197 &1628 &$-0.81 \pm 0.26$ &$4.0 \pm 1.2$ &\nodata\\
J115755.47+001704.0 &0.261 &1762 &$-0.24 \pm 0.29$ &$2.6 \pm 0.4$ &\nodata\\
J115832.81+005139.2 &0.591 &1035 &$-0.55 \pm 0.18$ &$3.2 \pm 0.4$ &\nodata\\
J121415.17+005511.4 &0.396 &1981 &\nodata &\nodata &\nodata\\
J122102.95-000733.7 &0.366 &517 &\nodata &\nodata &\nodata\\
J122432.40-002731.4 &0.157 &1308 &$-0.48 \pm 0.26$ &$3.0 \pm 0.5$ &\nodata\\
J124519.73-005230.4 &0.221 &1730 &\nodata &\nodata &\nodata\\
J125337.36-004809.6 &0.427 &1416 &$-0.35 \pm 0.39$ &$2.6 \pm 0.7$ &\nodata\\
J125943.59+010255.1 &0.394 &1459 &\nodata &\nodata &\nodata\\
J130023.22-005429.8 &0.122 &1018 &$-0.40 \pm 0.16$ &$2.6 \pm 0.3$ &\nodata\\
J130707.71-002542.9 &0.450 &1475 &\nodata &\nodata &\nodata\\
J130855.18+004504.1 &0.429 &1851 &\nodata &\nodata &\nodata\\
J131108.48+003151.8 &0.429 &1642 &\nodata &\nodata &\nodata\\
J132231.13-001124.5 &0.173 &1861 &$-0.78 \pm 0.45$ &$3.7 \pm 1.8$ &\nodata\\
J133031.41-002818.8 &0.240 &1216 &\nodata &\nodata &\nodata\\
J133741.76-005548.2 &0.279 &873 &\nodata &\nodata &\nodata\\
J135908.01+002732.0 &0.257 &1282 &\nodata &\nodata &\nodata\\
J141234.68-003500.0 &0.127 &1098 &\nodata &\nodata &\nodata\\
J141519.50-003021.6 &0.135 &1186 &$-0.46 \pm 0.18$ &$3.5 \pm 0.4$ &\nodata\\
J141820.33-005953.7 &0.254 &831 &\nodata &\nodata &\nodata\\
J142441.21-000727.1 &0.318 &1201 &$-0.58 \pm 0.17$ &$3.7 \pm 0.4$ &\nodata\\
J143030.22-001115.1 &0.103 &1744 &\nodata &\nodata &\nodata\\
J143230.99-005228.9 &0.362 &1559 &\nodata &\nodata &\nodata\\
J143624.82-002905.3 &0.325 &1857 &$-0.54 \pm 0.33$ &$3.7 \pm 0.8$ &\nodata\\
J144735.25-003230.5 &0.217 &1105 &\nodata &\nodata &\nodata\\
J144913.51+002406.9 &0.441 &944 &\nodata &\nodata &\nodata\\
J144932.70+002236.3 &0.081 &1072 &\nodata &\nodata &\nodata\\
J145123.02-000625.9 &0.139 &1122 &\nodata &\nodata &\nodata\\
J145437.84-003706.6 &0.576 &1328 &\nodata &\nodata &\nodata\\
J150629.23+003543.2 &0.370 &1861 &\nodata &\nodata &\nodata\\
J151312.41+001937.5 &0.159 &1697 &\nodata &\nodata &\nodata\\
J151956.57+001614.6 &0.115 &1716 &$0.40 \pm 0.34$ &$2.6 \pm 0.6$ &\nodata\\
J153243.67-004342.5 &0.309 &1877 &\nodata &\nodata &\nodata\\
J153911.17+002600.8 &0.265 &539 &$0.46 \pm 0.45$ &$2.8 \pm 0.9$ &\nodata\\
J164907.64+642422.3 &0.184 &759 &\nodata &\nodata &\nodata\\
J165022.88+642136.1 &0.407 &1152 &\nodata &\nodata &\nodata\\
J165338.69+634010.7 &0.279 &1848 &$-0.83 \pm 0.11$ &$4.4 \pm 0.6$ &\nodata\\
J165537.78+624739.0 &0.597 &1271 &\nodata &\nodata &\nodata\\
J165633.87+641043.7 &0.272 &1139 &\nodata &\nodata &\nodata\\
J165658.38+630051.1 &0.169 &1466 &$-0.35 \pm 0.11$ &$3.0 \pm 0.2$ &\nodata\\
J165905.45+633923.6 &0.368 &1359 &\nodata &\nodata &\nodata\\
J170546.91+631059.1 &0.119 &1657 &\nodata &\nodata &\nodata\\
J170812.29+601512.6 &0.145 &1094 &$-0.16 \pm 0.13$ &$2.6 \pm 0.2$ &\nodata\\
J170956.02+573225.5 &0.522 &1329 &$-0.36 \pm 0.16$ &$3.0 \pm 0.3$ &\nodata\\
J171033.21+584456.8 &0.281 &652 &\nodata &\nodata &\nodata\\
J171207.44+584754.5 &0.269 &1708 &$-0.34 \pm 0.10$ &$2.9 \pm 0.2$ &\nodata\\
J171540.92+560655.0 &0.297 &1752 &$-0.56 \pm 0.21$ &$3.5 \pm 0.5$ &\nodata\\
J171829.01+573422.4 &0.101 &1322 &$0.16 \pm 0.17$ &$2.3 \pm 0.3$ &\nodata\\
J172007.96+561710.7 &0.389 &1221 &$-0.47 \pm 0.20$ &$3.4 \pm 0.4$ &\nodata\\
J172206.04+565451.6 &0.426 &1579 &$-0.07 \pm 0.10$ &$2.8 \pm 0.2$ &\nodata\\
J172756.86+581206.0 &0.414 &1742 &$0.65 \pm 0.22$ &$1.5 \pm 0.5$ &\nodata\\
J172800.67+545302.8 &0.246 &1583 &$-0.06 \pm 0.19$ &$2.8 \pm 0.3$ &\nodata\\
J172823.61+630933.9 &0.439 &1750 &$0.61 \pm 0.29$ &$1.6 \pm 0.6$ &\nodata\\
J173404.85+542355.1 &0.685 &1163 &\nodata &\nodata &\nodata\\
J173721.14+550321.7 &0.333 &1256 &$0.11 \pm 0.28$ &$2.7 \pm 0.4$ &\nodata\\
J232525.53+001136.9 &0.491 &1921 &\nodata &\nodata &\nodata\\
J233032.95+000026.4 &0.123 &956 &\nodata &\nodata &\nodata\\
J233149.49+000719.5 &0.367 &1708 &\nodata &\nodata &\nodata\\
J233853.83+004812.4 &0.170 &1011 &\nodata &\nodata &\nodata\\
J234050.53+010635.6 &0.358 &729 &\nodata &\nodata &\nodata\\
J234141.50-003806.7 &0.319 &1871 &$0.00 \pm 0.38$ &$2.9 \pm 0.6$ &\nodata\\
J234150.81-004329.2 &0.251 &1817 &$-0.32 \pm 0.31$ &$3.4 \pm 0.5$ &\nodata\\
J234216.74+000224.1 &0.185 &917 &\nodata &\nodata &\nodata\\
J234229.46-004731.6 &0.316 &1857 &\nodata &\nodata &\nodata\\
J234725.30-010643.7 &0.182 &1667 &$0.52 \pm 0.25$ &$2.0 \pm 0.5$ &\nodata\\
\enddata

\tablenotetext{a}{Format: SDSS JHHMMSS.ss$\pm$DDMMSS.s}

\tablecomments{(1) Listed as a NLS1 in \citet{vv01}; (2) Previously classified as a Seyfert 1 by \citet{puch92}; (3) Listed as a Seyfert 1 in NED, due to reference in \citet{vv01}; (4) Previously listed as a NLS1 in VVG01; (5) Listed as a Seyfert galaxy in \citet{lcb92}}

\end{deluxetable}

\end{document}